\documentclass[aps,preprint,prb]{revtex4}

\usepackage[T1]{fontenc}
\usepackage{epsfig}
\usepackage{dcolumn}
\usepackage{bm}

\draft
\begin{document}

\title{Two--dimensional clusters of liquid $^4$He}
\author{A. Sarsa}
\address{Depto. de F\'{\i}sica Moderna, Universidad de Granada,
%\affiliation{Depto. de F\'{\i}sica Moderna, Universidad de Granada,
E-18071 Granada, Spain \\
International School for Advanced Studies, SISSA, 
Via Beirut 2/4, I-34014, Trieste, Italy}
\author{J. Mur-Petit, A. Polls}
\address{Dept.\ d'Estructura i Constituents de la Mat\`eria,
%\affiliation{Dept.\ d'Estructura i Constituents de la Mat\`eria,
Universitat de Barcelona, Diagonal, 647, E-08028 Barcelona, Spain}
\author{J. Navarro}
\address{IFIC, CSIC-Universitat de Val\`encia, Apdo. 20285,
%\affiliation{IFIC, CSIC-Universitat de Val\`encia, Apdo. 20285,
E-46071 Val\`encia, Spain}

\date{\today}

\begin{abstract}
The binding energies of two--dimensional clusters (puddles) of $^4$He are
calculated in the framework of the diffusion Monte Carlo method. The
results are well fitted by a mass formula in powers of $x=N^{-1/2}$,
where $N$ is the number of particles. The analysis of the mass formula allows
for the extraction of the line tension, which turns out to be 0.121 K/\AA. 
Sizes and density profiles of the puddles are also reported.
\end{abstract}

\pacs{36.40.-c 61.46.+w 67.70.+n}

\maketitle

\section{Introduction}
In recent years, a great deal of work has been devoted to study
quantum liquids in restricted geometries.\cite{kro02}  One important
feature of these systems is that their internal structure becomes more
easily observable than in bulk liquids due to the restricted motion of
the particles in the confining potential.  Among these systems the
study of quantum films has received particular attention. They consist
of liquid helium adsorbed to a more-or-less attractive flat surface. In
1973 M. Bretz {\it et al.}\cite{bre73} observed for the first time 
the adsorption of $^4$He onto the basal plane of graphite. In the last
few years, adsorption properties of helium on different substrates
such as carbon, alkali and alkaline-earth flat surfaces, carbon
nanotubes and aerogels have become a fertile topic of research.

The structure and growth of thin films of $^4$He adsorbed to a substrate was 
studied by Clements {\it et al.}\cite{cle93} employing the optimized
hypernetted-chain Euler-Lagrange theory with realistic atom-atom
interactions. It turns out that films with low surface coverages,
where all atoms cover the surface with a thickness corresponding to a
single atom, can be approximated reasonably well by a 2D model. In
connection with these systems, an interesting question naturally
arises as how physics depends on the dimensionality of the space.

The homogeneous 2D liquid has been studied using different theoretical
methods, such as molecular dynamics\cite{cam71} and quantum Monte
Carlo simulations either Green's Function\cite{whi88} or
diffusion\cite{gio96} techniques. The inhomogeneous case was studied
by Krishnamachari and Chester,\cite{kri99} who used a shadow variational 
wave function to describe 2D puddles of liquid $^4$He. In this work we report
energies and density profiles of  puddles calculated within the diffusion
Monte Carlo (DMC) method. Our main objective is to give an accurate
estimate of the line energy or the line tension of the 2D liquid $^4$He. As
atom-atom interaction we have used the revised version of the Aziz
potential dubbed as HFD-B(HE).\cite{azi87} This potential has been
used to study ground--state properties of 3D bulk $^4$He and 
$^3$He,\cite{bor94,cas00} within the DMC framework, and it has proved
to accurately reproduce the ground--state properties of both liquids at
zero temperature.

The trial wave function used for the importance sampling in the DMC
calculation is introduced in Section II, where the variational Monte
Carlo results for this wave function are also reported.
A brief explanation of the DMC techniques  used in the present paper
is presented in Section III.
Section IV contains the DMC results and their analysis in terms of a
mass formula in 2D. The line tension is extracted from this mass
formula. Properties characterizing the puddles, such as the density
profiles, are discussed in Section V.
Finally, the main conclusions are summarized in Section VI.

\section{Variational ground--state energies}
To study a system of $N$ $^4$He atoms in two dimensions 
we start from the following trial wave function 
\begin{equation}
  \label{eq:wf}
  \Phi_T({\bf R}) = \prod_{i<j} \exp \left[ 
    - \frac{1}{2} \left(\frac{b}{r_{ij}}\right)^{\nu}
    - \frac{\alpha^2}{2N} r^2_{ij} \right] \, ,
\end{equation}
written in the same way as in the 3D case.\cite{panda86}
The coordinate ${\bf R}$ indicates the set of coordinates
of all the particles $\{ {\bf r}_1,{\bf r}_2,...,{\bf r}_N \}$,
 while $r_{ij}$ stands for the interparticle distance, 
$r_{ij}=|{\bf r}_j-{\bf r}_i|$.
The trial wave function  contains the simple McMillan form\cite{mcm65} to
deal with the very short-range part of the interaction, and the
translationally invariant part of a harmonic oscillator (HO) wave function
with parameter $\alpha$, to roughly confine the system. 

In our calculations the value $\hbar^2/m_4 = 12.1194$~K~\AA$^2$ has been 
employed for the atom mass and the parameters $b$ and $\nu$ have been fixed to 
the values 3.00~\AA\ and 5, respectively, the same values as in 3D 
calculations. The variational search has thus been restricted to the HO 
parameter $\alpha$, whose optimal value is given in Table~\ref{var_ene}. 
The expectation value of the Hamiltonian, as well as the separate
contributions of kinetic and potential energies are given in the same Table
for puddles with $N$ atoms. It can be seen that the total energy results from
an important cancellation between kinetic and potential energies, which is in
fact larger than in the 3D case. Let us recall that in 3D bulk, the energy per
particle results from adding $\approx$  14~K of kinetic energy with 
$\approx -21$~K of potential energy.  In 2D, both kinetic and potential
contributions are very close to each other, which makes the calculation very
delicate.

In the last column of Table~\ref{var_ene}  the VMC results of 
Krishnamachari and Chester\cite{kri99} are reported. As compared with their 
results, our calculations provide smaller binding energies, in spite of the 
fact that the interaction used in Ref.~\onlinecite{kri99} is an older version 
of the Aziz potential, which tends to underbind the systems. In fact, the 
shadow wave function used in Ref.~\onlinecite{kri99} contains more elaborated 
correlations not present in our simple trial wave function.  The VMC energy
for the bulk system corresponds to the saturation density, 
$\rho_0= 0.04344$ \AA$^{-2}$, taken from the DMC calculation of 
Ref.~\onlinecite{gio96}.

We have also performed calculations using a different trial wave function, 
replacing the translationally invariant gaussian part by an exponential one, 
i.e.
\begin{equation}
  \label{eq:wf-expo}
  \Phi_T({\bf R}) = \prod_{i<j} \exp \left[ 
    - \frac{1}{2} \left(\frac{b}{r_{ij}}\right)^{\nu}
    - \frac{\alpha}{2} r_{ij} \right],
\end{equation}
expecting that this larger tail in the wave function will result in more
binding. Actually, we do not find significant differences for small values of 
$N$. For instance, in the case $N=8$, using the same values for $b$ and $\nu$
as before, we get $E/N=-0.2178(5)$ K, $T/N=1.266(2)$ K,
$V/N=-1.484(2)$ K for $\alpha=0.035$ \AA$^{-1}$. When the values of
$b$, $\nu$ and $\alpha$ are optimized, we obtain a slightly larger
binding energy, $E/N=-0.2267(8)$ K for $b=3.04$ \AA, $\nu=5.0$ and
$\alpha=0.035$ \AA$^{-1}$. For greater values of $N$, the gaussian
ansatz tends to provide more binding than the exponential. For
example, with the exponential ansatz, for $N=16$ we get $E/N=-0.1816(7)$
K for $\alpha=0.023$ \AA$^{-1}$, and optimizing the different
parameters one gets $E/N=-0.2514(6)$ K, with $b=3.04$ \AA. In conclusion,
the gaussian wave function seems appropriate to be used as importance 
sampling in the DMC calculations.

\section{Diffusion Monte Carlo ground--state energies}
Quantum Monte Carlo (QMC) methods provide the exact ground--state energy of a
boson system, except for statistical errors. These techniques solve
numerically the Schr\"odinger equation by means of a statistical
simulation. They have been  
widely described in the literature; hence we briefly recall  here the main 
ideas, referring the reader to, for example, Ref.~\onlinecite{hlr94} for a
more detailed description on QMC techniques. In this work we use the diffusion
Monte Carlo (DMC) method to solve the Schr\"odinger equation in imaginary time 
($\tau = i t$) for the function
\begin{equation}
f({\bf R},\tau) = \Phi_T({\bf R}) \Psi({\bf R}, \tau)
\end{equation}
where ${\bf R}$ represents all the particle coordinates and is usually called
``the walker'', $\Psi({\bf R}, \tau)$ is the wave function of the system, and
$\Phi_T({\bf R})$ is the previously determined trial wave function (Sect. II),
used here as  importance sampling. It is convenient to write the solution of
the time-dependent Schr\"odinger equation in the following form
\begin{equation}
f({\bf R},\tau+\Delta \tau) = \int d{\bf R'} G({\bf R},{\bf R'}, \Delta \tau)
f({\bf R'},\tau)
\label{schroint}
\end{equation}
where $G$ is the time-dependent Green's function and 
is formally written as 
\begin{equation}
  G({\bf R},{\bf R'};\tau)=
  \langle {\bf R} \mid e^{-H\,\tau} \mid {\bf R'} \rangle
\label{lage}
\end{equation}
where $H$ is the Hamiltonian of the system. The function 
$G({\bf R},{\bf R'};\tau)$
represents the amplitude probability for the transition from an initial
state ${\bf R'}$ to a final one ${\bf R}$ in a time $\tau$. In the limit
$\tau \rightarrow \infty$, Eq.~(\ref{schroint}) gives the exact ground--state
wave function. Thus, knowing $G$ for infinitesimal time steps $\Delta
\tau$, the asymptotic solution for large times $f({\bf R}, \tau \to \infty)$
can be obtained by solving iteratively the above equation. To this end, the 
exponential entering Eq.~(\ref{lage}) is approximated to some fixed order in 
$\Delta \tau$. 
Both first and second order\cite{bor94} propagators have been implemented in
the present work and both of them provide the same extrapolated energy, within
statistical errors, using the trial function $\Phi_T$ of Eq.~(\ref{eq:wf}) as
guiding function. 
Our simulations have been carried out with a population of typically 400
walkers. As usual, some runs are first done to establish the asymptotic
region of the short time propagator, then several values of the time step
have been used, and finally a fit, either linear or quadratic, has been
carried out to obtain the extrapolated energy.
For example, for $N=16$ the time steps 0.0001, 0.0002, 0.0003, and 0.0004
have been used to perform the extrapolation. In general, the statistical
error is of the order of the time step error in our calculations.

In Table~\ref{dmc_ene} we present the results of our linear  DMC calculations 
of the total energy per particle for puddles containing $N$ atoms. 
We have also reported and reproduced the results of the binding energy per
particle of homogeneous 2D liquid $^4$He at the equilibrium density
$\rho_0^{DMC}=0.04344(2)$ \AA$^{-2}$; obtained in Ref.~\onlinecite{gio96},
where the same version of the Aziz potential was used. For this case, the
simulations have been carried out for a system of 64 atoms with periodic
boundary conditions, for which the errors due to finite size effects are
smaller than the statistical ones\cite{whi88}.
We have also performed quadratic DMC calculations for some puddles, 
and found results which are compatible with the linear DMC ones within their 
error bars. For example, the quadratic algorithm provides 
$E_{quad}=-0.2612(2)$ K for $N=8$ and $-0.652(4)$ for $N=64$. 
As expected, the DMC results lower the corresponding energies 
obtained by VMC either with our simple variational wave function or with a 
shadow wave function,\cite{kri99} by up to $\sim 25 \%$ in the case of the 
bulk system. 
It is worth mentioning that the final DMC result for the energy does not 
depend on the trial wave function for a boson system like the studied
here, a fact that in the present case has been numerically checked for the
Gaussian and the exponential ansatzs, Eqs.~(\ref{eq:wf})
and~(\ref{eq:wf-expo}). Indeed, for boson systems the DMC method provides
exact ground--state energies, within statistical errors.

\section{Energy and line tension}

For a saturating self-bound system, the ground--state energy per
particle can be expanded in a series of powers of the variable
$N^{-1/D}$, where $N$ is the number of constituents and $D$ is the
dimensionality of the space. This is the well-known mass formula, 
which in the present case writes
\begin{equation}
  \label{eq:mass-f}
  E(N)/N = \varepsilon_{b} + \varepsilon_{l} x + \varepsilon_{c} x^2 + \cdots
\end{equation}
with $x=N^{-1/2}$. The two first coefficients of this expansion are
the bulk energy $\varepsilon_{b}$ and the line energy $\varepsilon_{l}$, out
of which the line tension $\lambda$ is defined by $2 \pi r_0 \lambda =
\varepsilon_{l}$. Here $r_0$ is the unit radius, defined as the radius of
a disk whose surface is equal to the inverse of the equilibrium
density of the 2D bulk liquid, i.e. $\rho_0 \pi r_0^2=1$.
Finally, $\varepsilon_c$ is the so-called curvature energy.

\begin{figure}[t]
\includegraphics*[angle=-90, width=12cm]{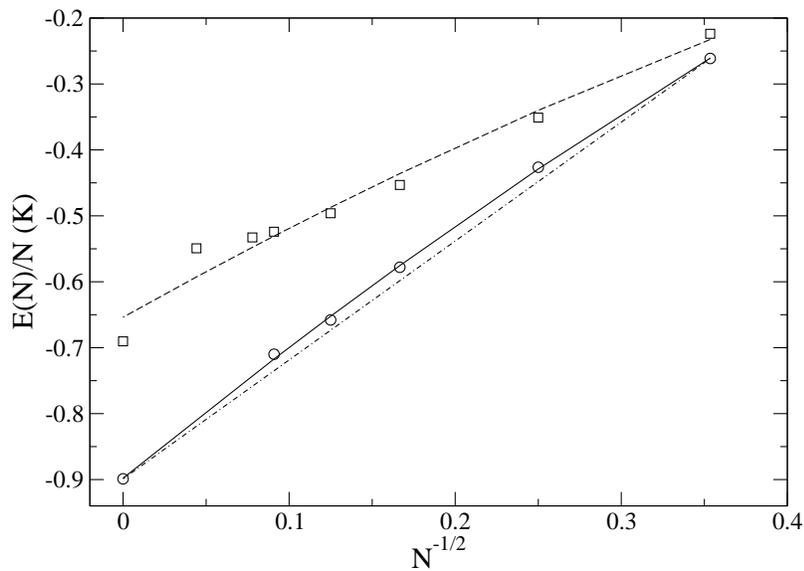}
\caption{Energies per particle (in K) of $N$-atom puddles as a 
function of $N^{-1/2}$, obtained from our VMC (squares) and DMC (circles) 
calculations. The interaction used is Aziz HFD-B(HE). Dashed and solid lines 
correspond to a least square fit to these energies. The dot-dashed line is a 
straight line between the $N=8$ and bulk DMC values.
\label{energies}}
\end{figure}

Our calculated ground--state energies (Tables~\ref{var_ene} and
~\ref{dmc_ene}) are plotted in Fig.~\ref{energies} as a function of 
$N^{-1/2}$. One can see that the differences between our VMC and DMC energies
increase with the number of atoms in the puddle. This clearly mirrors the
simplicity of the trial wave function, which could  be improved by including,
for example, three-body correlations. Nevertheless this trial function is
adequate for the importance sampling in the DMC calculation.

We have fitted these energies to a parabolic mass formula like 
Eq.~(\ref{eq:mass-f}). The coefficients of the fit are given in 
Table~\ref{coefs}, together with the deduced line tension. 
Notice that the coefficient $\varepsilon_b$ is identical, 
within statistical errors, to the 
bulk energy per particle of Table~\ref{dmc_ene}. 
In fact, the $\chi^2$ of the fit is very small, $\chi^2=5.7\times10^{-6}$.
Regarding the line tension, and despite using a different version of the Aziz
potential and a different trial function, we notice that our VMC estimate is
rather close to the one reported in Ref.~\onlinecite{kri99}, $\lambda=0.07$
K/\AA. However, both VMC results are remarkably different from the DMC line
tension, $\lambda=0.121$ K/\AA. 
 
To stress the curvature effect we have also plotted in the figure a straight 
line between the $N=8$ and bulk DMC values. A linear fit of the DMC energies 
gives coefficients $\varepsilon_b=-0.885$~K and $\varepsilon_l=1.80$~K,
which are appreciably different from the previous ones. The  bulk energy 
extrapolated from this linear  fit differs from the directly calculated value, 
and the corresponding line energy is closer to the variational one, thus 
giving a bad estimation for the linear tension. 
In all cases, the line energy coefficient is approximately minus twice 
the volume energy term, similarly to the 3D case,\cite{panda86} and 
therefore one expects curvature effects to be important. 
In both VMC and DMC cases the extracted $\varepsilon_c$ is negative, 
i.e. the binding energy is a convex function of $x$ as it also happens for the
3D clusters\cite{panda86}. This is in contrast with the value of
$\varepsilon_c$ reported in Ref.~\onlinecite{kri99} which was positive
but rather smaller in absolute value and with larger error bars.
Actually, as argued in Ref.~\onlinecite{panda86} for 3D clusters, 
one would expect the curvature correction to the energy of a circular 
2D cluster to be positive. Therefore, one should take the extracted 
value for $\varepsilon_c$ with certain caution and not to emphasize 
its physical significance.
However, it turns out that the value and sign of $\varepsilon_c$ are 
stable against different possible fits, e.g., changing the number of 
points to build the fit, or using a cubic mass formula.
In any case the two first coefficients $\varepsilon_b,\varepsilon_l$
are quite robust against all performed fits. As an illustration, 
if one takes out the bulk point, the predicted bulk energy per 
particle and surface tension are equal to the reported ones within 1\% 
and 5\% respectively. 
%e_b=-0.898+/-0.02, e_L=2.05+/-0.02, e_c=-0.71+/-0.03 => l=0.121+/-0.001 K/A
%e_b=-0.889+/-0.04, e_l=1.96+/-0.03, e_c=-0.52+/-0.05 => l=0.115+/-0.002 K/A
Therefore, the extracted line tension should be reliable, as it also
happens for 3D clusters\cite{panda86}.

\section{Density profiles}

The calculation of observables given by operators that do not commute
with the Hamiltonian poses new problem to the DMC method. After
convergence, the walkers are distributed according to the so-called
mixed probability distribution given by the product of the exact and
the trial wave functions. Therefore averaging the local values of the
operator does not give the exact expectation value unless the operator
commutes with the propagator.  The result obtained by straightforward
averaging is the mixed estimator which is of first order error in the
trial wave function. Several options have been proposed in the
literature in order to obtain unbiased (trial function independent and
exact) values. 
In this work we have used the so called {\it forward} or {\it future walking}
technique\cite{hlr94} to calculate unbiased, also called pure, density
profiles. The key ingredient to correct the mixed estimator is to 
include as a weight in the sampling the quotient 
$\Phi_{\rm exact}({\bf R})/\Phi_{\rm trial}({\bf R})$ for each walker, 
given by the asymptotic number of walkers. Several algorithms have been
proposed in order to compute this quantity. In this work we use the algorithm
developed in Ref.~\onlinecite{cas95} that constitutes a simple and efficient
implementation of the future walking method.

\begin{figure}[t]
\includegraphics*[width=12cm]{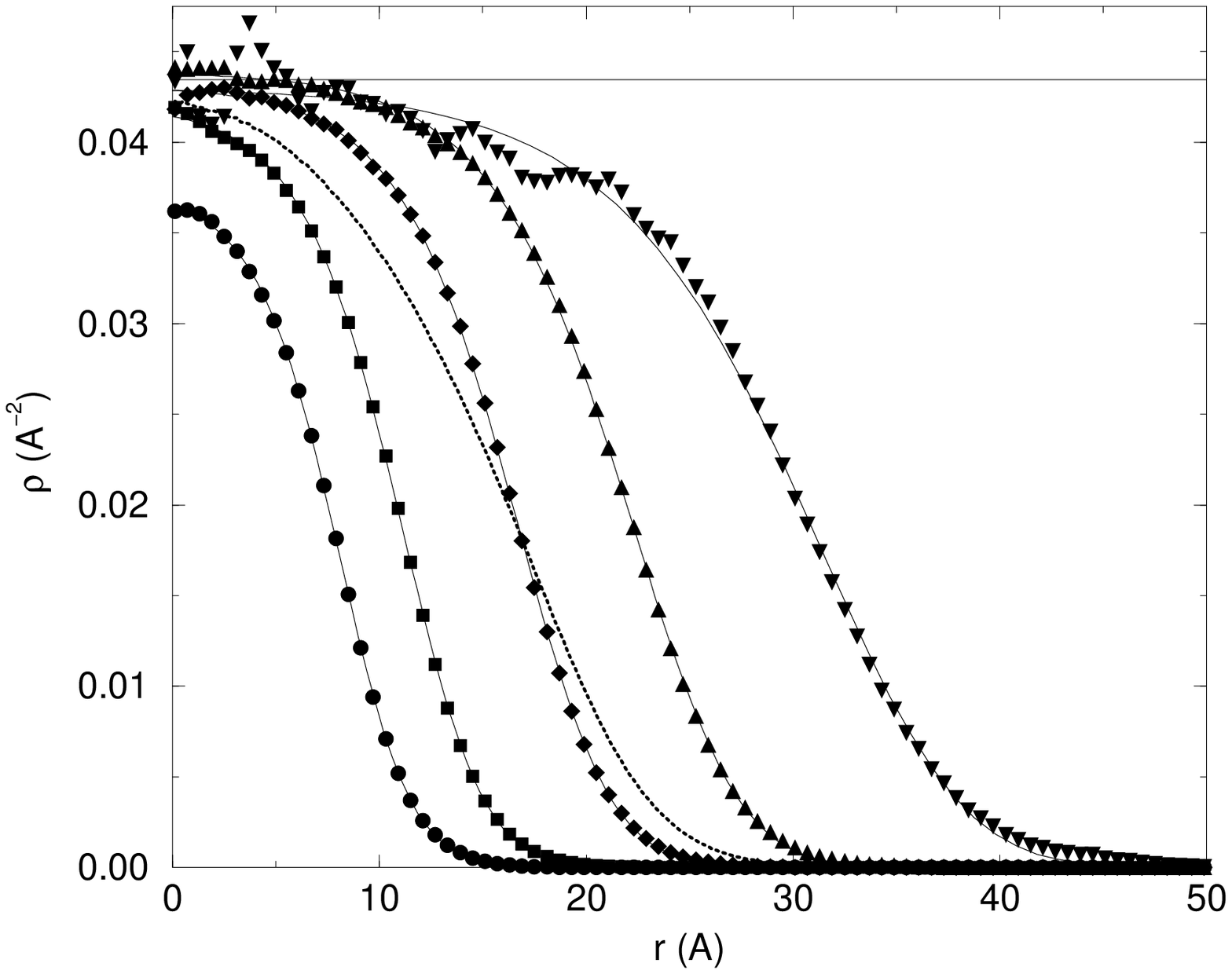}
\caption{Density profiles for $^4$He puddles with various number of
atoms, $N=8$ (circles), 16 (squares), 36 (diamonds), 64 (triangles up) 
and 121 (triangles down), obtained from our pure estimators for the
linear DMC calculations. 
The solid horizontal line indicates the saturation density of the 
homogeneous system. 
The dotted line is the VMC profile for $N=36$ with a Gaussian trial function.
The figure also contains the fits to the data provided by a generalized 
Fermi function, as explained in the text.} 
\label{profiles}
\end{figure}

The pure DMC estimates of the density profiles for several puddles are plotted
in Fig.~\ref{profiles}. The figure also contains an horizontal line which
indicates the saturation density ($\rho_0^{DMC}$) of the homogeneuos system. 
For the puddle containing 36 atoms, the VMC profile  obtained from a
Gaussian ansatz (Eq.~(\ref{eq:wf})) is also shown for comparison as a dotted
line. As one can appreciate in the figure, the process of optimization implied
by the DMC method changes the profile reducing its thickness, i.e., producing
a sharper surface. 
It can be seen that for the smaller clusters the central density is below
$\rho_0$, while for the larger values of $N$ shown in the figure the central
density is above $\rho_0$, indicating a leptodermous behaviour. One expects
that, increasing the number of particles, the central density will approach
$\rho_0$ from above, as in the 3D case.\cite{str87,chi92} 
It is worth noticing the oscillating behaviour in the interior part of the
density profile for $N=121$. It is difficult however to decide whether these
oscillations are genuine or are simply due to a poor statistics in evaluating
the pure estimator. Unfortunately, to discard this last option would require
an exceedingly long computing time, within the scheme of this work.

The solid lines plotted in Fig.~\ref{profiles} are fits to our DMC densities
provided by a generalized Fermi profile of the form:
\begin{equation}
\rho(r)= \frac{\rho_f}{\left( 1+\exp{\left(\frac{r-R}{c}\right)} \right)^{\nu}}
\end{equation} 
The parameters defining the Fermi profile are given in Table~\ref{fermi}
together with the thickness $t$ and the root mean square (rms) radius
$\langle r^2 \rangle^{1/2}$. We have checked that the rms radius calculated
within the DMC code and the one derived form the fit agree to better than
0.5\%, except for the $N=121$ case, where the difference is 1\%.
The rms radius grows with the number of particles as $N^{1/2}$, as
expected. Therefore it grows faster than in 3D, in which case grows as
$N^{1/3}$. This behaviour allows for an alternative determination of the
saturation density by performing a linear fit to the  relation
\begin{equation}
\sqrt{\langle r^2 \rangle^{1/2}} = \sqrt{ \frac {N}{2 \pi \rho_0}}
\end{equation}
The extracted value of $\rho_0$ from the mean square radius reported in
Table~\ref{fermi} is 0.043 \AA$^{-2}$, in good agreement with the
determination from the calculation for the homogeneous system.

In the interval of $N$ considered, the thickness $t$, defined as the distance
over which the density falls from 0.9 of its value at origin to 0.1, is 
continuously increasing.  However, as the finite value of the thickness
for the semiinfinite system should define the asymptotic value of $t$, one 
expects that for larger puddles the thickness will probably have a maximum and 
smoothly approach this asymptotic value, as happens in the 3D
case.\cite{str87} 
Finally, one also observes the asymmetric character of the density profiles 
with respect to the point at which the density falls at half its value
at the origin. This can be appreciated by looking at the value of $\nu$, which 
grows with $N$, and also to the increasing difference between the quantities 
$R$ and $\langle r^2 \rangle^{1/2}$.

\section{Summary and conclusions}

In this work we have considered strictly two--dimensional systems  of liquid
$^4$He, which  are of course an idealization of a real quantum film.
They are  nevertheless interesting  because their study can enlighten 
 the underlying structure of  real quasi-2D systems. Of course, in the
latter case, one has to take also into account the interaction with the
substrate, which basically provides a global attractive potential.
In the ideal 2D case, the suppression of the wave function component in the 
third dimension, produces an increment of the global repulsion between atoms, 
resulting in a smaller binding energy per particle, and a decrease of the 
equilibrium density.\cite{apa}

The binding energies of two--dimensional $^4$He clusters, calculated by means 
of  a diffusion Monte Carlo method,  are well fitted by a mass formula in 
powers of $x=N^{-1/2}$. The analysis of the mass formula provides the main 
result of this paper, namely the value of the line tension 
$\lambda =0.121 $ K/\AA, which significantly differs from the one obtained 
from a similar analysis of VMC data and the one previously reported in the
literature\cite{kri99}. The quadratic term of the mass formula cannot be
neglected and results in a negative value of the curvature energy as in the 3D
case.\cite{panda86,chi92} However, the studied clusters may be too small to
give physical significance to this result.

The density profiles obtained with the pure estimator have been fitted by a
generalized Fermi function, and the behaviour of the rms radius and the
thickness as well as the asymmetry character of the profile as a function of
$N$ have been discussed. However, calculations for larger puddles, which are
out of the scope of the present paper due to limitations in computing time,
would be necessary to describe the complete $N$-dependence of the density
profiles.
 
\acknowledgments
Fruitful discussions with X. Viñas are gratefully acknowledged.
This work has been supported by DGICYT (Spain) project PB98-1247,
DGI (Spain) grants BFM2001-0262 and BFM2002-00200, Generalitat de Val\`encia
grant GV01-216, Ge\-ne\-ra\-litat de Catalunya project 2001SGR00064, and
MURST (Italy) grant MIUR-2001025498. J. M. P. acknowledges a
fellowship from the Ge\-ne\-ra\-litat de Catalunya.

%%%%%%%%% REFERENCES %%%%%%%%%%%%

%%%%%%%%% TABLES %%%%%%%%%%%%

\pagebreak

\begin{table}[t]
\begin{center}
\caption{Variational results for the ground--state energy per particle $E/N$
  of 2D $^4$He puddles of various cluster sizes. The confining HO parameter
  $\alpha$ is given in \AA$^{-1}$ and all energies are in K. The expectation
  values of the kinetic and the potential energies are also displayed. The
  column labelled KC refers to the VMC results of Ref.~\onlinecite{kri99}.}
\label{var_ene}
\vspace{0.3cm}
\begin{tabular}{cccccc}
\hline
$N$      & $\alpha$ &  $E/N$     & $T/N$     & $V/N$      & KC  \\
\hline
8        & 0.1565   & -0.2239(2) & 1.3003(6) & -1.5242(5) & --- \\
16       & 0.129    & -0.3510(2) & 1.7354(6) & -2.0864(5) & -0.380(8)\\
36       & 0.094    & -0.4532(4) & 2.031(3)  & -2.484(3)  & -0.471(7)\\
64       & 0.073    & -0.4961(7) & 2.159(2)  & -2.655(2)  & -0.528(5)\\
121      & 0.054    & -0.5241(6) & 2.223(2)  & -2.747(2)  & -0.570(7)\\
165      & 0.047    & -0.5328(3) & 2.289(1)  & -2.822(1)  & -0.602(7)\\
512      & 0.0266   & -0.5493(5) & 2.282(3)  & -2.831(3)  & -0.621(2)\\
$\infty$ & 0.0000   & -0.6904(8) & 4.312(2)  & -5.003(1)  & --- \\
\hline
\end{tabular}
\end{center}
\end{table}

\pagebreak

\begin{table}[t]
\begin{center}
\caption{Energy per particle (in K) for 2D $^4$He puddles for various 
  cluster sizes obtained with the DMC algorithm.}
\label{dmc_ene}
\vspace{0.3cm}
\begin{tabular}{c|cccccc}
\hline
$N$  & 8         & 16         & 36        & 64        & 121       &$\infty$  \\
$E/N$&-0.2613(4) & -0.4263(4) & -0.578(2) & -0.658(4) & -0.710(2) & -0.899(2)\\
%$N$      & {\rm $E/N$} \\ %& {\rm quadratic} \\
%\hline
%8        & -0.2613(4)   \\ %& -0.2612(2)      \\
%16       & -0.4263(4)   \\ %& -0.426(1)       \\
%36       & -0.578(2)    \\ %& -0.575(3)       \\
%64       & -0.658(4)    \\ % & -0.652(4)       \\
%121      & -0.710(2)    \\ %&                 \\
%$\infty$ & -0.899(2)    \\ %& -0.8971(6)  \\
\hline
\end{tabular}
\end{center}
\end{table}

\pagebreak

\begin{table}[t]
\begin{center}
\caption{Coefficients (in K) of a parabolic fit of the mass formula, 
as given in Eq.~(\ref{eq:mass-f}). The last column displays the deduced
line tension (in K~\AA$^{-1}$).}
\label{coefs}
\vspace{0.3cm}
\begin{tabular}{c|ccc|c}
\hline
 & $\varepsilon_b$      & $\varepsilon_l$ & $\varepsilon_c$ & $\lambda$ \\
\hline
VMC & -0.654(1) & 1.41(1) & -0.62(2) & 0.083(1) \\
DMC & -0.898(2) & 2.05(2) & -0.71(3) & 0.121(1) \\
%Quadratic DMC & -0.898 & 2.05 & -0.71 & 0.121 \\
\hline
\end{tabular}
\end{center}
\end{table}

\pagebreak

\begin{table}[t]
\begin{center}
\caption{Parameters of a Fermi-profile fit to the density profiles. All
  lengths are in \AA\ and $\rho_f$ is in \AA$^{-2}$. The parameter $\nu$ is
  adimensional.}
\label{fermi}
\vspace{0.3cm}
\begin{tabular}{c|cccc|cc}
\hline
$N$ & $\rho_f$ & $R$ & $c$   & $\nu$ & $t$   & $\langle r^2 \rangle^{1/2}$ \\ 
\hline
8   & 0.03740 & 9.308 & 2.156 & 1.739 & 8.166 &  7.20\\% DMC: & 7.23\\
16  & 0.04204 & 13.38 & 2.656 & 2.284 & 9.580 &  9.18\\% DMC: & 9.22\\
36  & 0.04305 & 19.47 & 3.104 & 2.400 & 11.11 & 12.91\\% DMC: & 12.94\\
64  & 0.04386 & 26.68 & 3.783 & 3.111 & 13.09 & 16.68\\% DMC: & 16.76\\
121 & 0.04304 & 40.09 & 5.566 & 4.714 & 18.52 & 23.15\\% DMC: & 23.39\\
\hline
\end{tabular}
\end{center}
\end{table}

%%%%%%%%% FIGURES %%%%%%%%%%%%

\pagebreak

\begin{thebibliography}{99}

\bibitem{kro02} See, e.g., E. Krotscheck and J. Navarro (Eds.), {\it
Microscopic Approaches to Quantum Liquids in Confined Geometries},
Series on Advances in Quantum Many-Body Theories, Vol. 4 (World
Scientific, Singapore,2002).

\bibitem{bre73} M. Bretz, J. G. Dash, D.C. Hickernell, E. O. McLean
and O. E.  Vilches, Phys. Rev. A {\bf 8}, 1589 (1973).

\bibitem{cle93} B. E. Clements, J. L. Epstein, E. Krotscheck and M.
Saarela, Phys. Rev. B {\bf 48}, 7450 (1993).

\bibitem{cam71} C. E. Campbell and M. Schick, Phys. Rev. A {\bf 3},
691 (1971).

\bibitem{whi88} P. A. Whitlock, G. V. Chester and M. H. Kalos, Phys.
Rev.  B {\bf 38}, 2418 (1988).

\bibitem{gio96} S. Giorgini, J. Boronat and J. Casulleras, Phys. Rev.
B {\bf 54}, 6099 (1996).

\bibitem{kri99} B. Krishnamachari and G. V.  Chester, Phys. Rev. B
{\bf 59}, 8852 (1999).

\bibitem{azi87} R. A. Aziz, F. R. McCourt and C. C. K. Wong, Mol.
Phys. {\bf 61}, 1487 (1987).

\bibitem{bor94} J. Boronat and J. Casulleras, Phys. Rev. B {\bf 49},
8920 (1994).

\bibitem{cas00} J. Casulleras and J. Boronat, Phys.  Rev. Lett. 
{\bf 84}, 3121 (2000).

\bibitem{panda86} V. R. Pandharipande, S. C. Pieper and R. B.
Wiringa, Phys. Rev. B {\bf 34}, 4571 (1986).

\bibitem{mcm65} W. L. McMillan, Phys. Rev. {\bf 138}, A442 (1965).

\bibitem{hlr94} B. L. Hammond, W. A. Lester Jr. and P. J. Reynolds,
{\em Monte Carlo Methods in ab initio Quantum Chemistry}, Lecture and Course
Notes in Chemistry, Vol. 1 (World Scientific, Singapore, 1994).

\bibitem{cas95} J. Casulleras and J. Boronat, Phys. Rev. B {\bf 52}, 
3654 (1995). 

%\bibitem{sar02} A. Sarsa, J. Boronat and J. Casulleras, J. Chem. Phys.
%{\bf 116}, 5956 (2002).
%esta referencia ya no aparece citada

\bibitem{str87} S. Stringari and J. Treiner, J. Chem. Phys. {\bf 87}, 5021
(1987).

\bibitem{chi92} S. A. Chin and E. Krotscheck, Phys. Rev. B {\bf 45}, 852 (1992).

\bibitem{apa} See, e.g., V. Apaja and E. Krotscheck, chapter 5 in
  Ref.~\onlinecite{kro02}, and references therein.

\end{thebibliography}
\end{document}